\documentclass[twocolumn,showpacs,aps,prl,preprintnumbers,superscriptaddress]{revtex4}

\usepackage{graphicx}
\usepackage{dcolumn}
\usepackage{amsmath}
\usepackage{epsfig}

\long\def\inst#1{\par\nobreak\kern 4pt\nobreak
    {\itshape #1}\par\vskip 10pt plus 3pt minus 3pt}

\RequirePackage{xspace}

\usepackage{relsize}

\include{plb_symbols}
\def\pipi   {\ensuremath{\pi\pi}}
\def\pippim {\ensuremath{\pip\pim}}

\def\Kpip   {\ensuremath{ (K\pi)^{+} }}
\def\mkpi   {\ensuremath{m_{K\pi}}}
\def\mpipi   {\ensuremath{m_{\pippim}}}

\def\calB      {\ensuremath{{\cal B}}\xspace}

\def\fL      {\ensuremath{ f_{L} }}

\def\etal {{\em et al.}}

\def\RhoKstar      {\ensuremath{\rho K^{\ast}}}   
\def\BtoRhoKstar   {\ensuremath{\B\to\RhoKstar}}  
\def\BtoRhopKstarz {\ensuremath{\Bp\to\rhop K^{\ast0}}} 

\def\fzKstarp        {\ensuremath{f_0(980)K^{\ast+}}}

\def\BtofzKstarp     {\ensuremath{\Bp\to\fzKstarp}}

\def\KstarpKp        {\ensuremath{K^{\ast+}\to\Kp\piz}}
\def\KstarpKz        {\ensuremath{K^{\ast+}\to\KS\pip}}

\def\RhozKstarp      {\ensuremath{\rho^0 K^{\ast+}}}

\def\BtoRhozKstarp   {\ensuremath{\Bp\to\RhozKstarp}}

\def\BtoDpiKz  {\ensuremath{\Bp\to\Dzb (\to\KS\pip\pim)\pip}}
\def\BtoDpiKp  {\ensuremath{\Bp\to\Dzb (\to\Kp\piz\pim)\pip}}

\def\fz   {\ensuremath{f_0(980)}\xspace}

\def\KstarII  {\ensuremath{K_{0}^{*}(1430)}\xspace}

\def\Bmeson  {\B\ meson}
\def\Bmesons {\B\ mesons}

\def\Dmeson  {\ensuremath{D} meson}
\def\Dmesons {\ensuremath{D} mesons}

\def\Bback   {\BB\ background}
\def\Bbacks  {\BB\ backgrounds}
\def\fl {\fL}

\newcommand{\onreslumi}  {\mbox{426\invfb}}
\newcommand{\offreslumi} {\mbox{44.4\invfb}}
\newcommand{\nbb}        {\mbox{$(467\pm5)\times 10^{6}$}}

\newcommand{\kzntot}     {\mbox{$7444$}}

\newcommand{\kznuds}     {\mbox{$5959\pm96$}}
\newcommand{\kznbb}      {\mbox{$1266\pm81$}}

\newcommand{\kzflstat}   {\mbox{$0.74\pm0.13$}}
\newcommand{\kzfl}       {\mbox{$\kzflstat\pm0.03$}}

\newcommand{\kznsig}     {\mbox{$85\pm24$}}  
\newcommand{\kzup}       {\mbox{$6.4$}} 
\newcommand{\kzsigfull}  {\mbox{$4.1$}} 
\newcommand{\kzbfsyst}   {\mbox{$0.5$}} 
\newcommand{\kzbf}       {\mbox{$4.6 \pm 1.2 \pm \kzbfsyst $}}

\newcommand{\fzkznsig}     {\mbox{$69\pm14$}}  
\newcommand{\fzkzsigfull}  {\mbox{$6.0$}} 
\newcommand{\fzkzbfsyst}   {\mbox{$0.3$}} 
\newcommand{\fzkzbf}       {\mbox{$3.6 \pm 0.7 \pm \fzkzbfsyst $}}

\newcommand{\kzeffavg}   {\mbox{$17.1$}}
\newcommand{\fzkzeffavg}   {\mbox{$17.9$}}

\newcommand{\kpntot}     {\mbox{$12867$}}

\newcommand{\kpnuds}     {\mbox{$10727 \pm 122$}}
\newcommand{\kpnbb}      {\mbox{$1451 \pm 129$}}

\newcommand{\kpflstat}   {\mbox{$0.94\pm0.27$}}
\newcommand{\kpfl}       {\mbox{$\kpflstat\pm0.03$}}

\newcommand{\kpnsig}     {\mbox{$67\pm31$}}  
\newcommand{\kpup}       {\mbox{$7.1$}} 
\newcommand{\kpsigfull}  {\mbox{$3.3$}} 
\newcommand{\kpbfsyst}   {\mbox{$0.5$}} 
\newcommand{\kpbf}       {\mbox{$4.4 \pm 2.0 \pm\kpbfsyst$}}

\newcommand{\fzkpnsig}     {\mbox{$91\pm20$}}  
\newcommand{\fzkpsigfull}  {\mbox{$6.8$}} 
\newcommand{\fzkpbfsyst}   {\mbox{$0.3$}} 
\newcommand{\fzkpbf}       {\mbox{$5.2 \pm 1.0 \pm \fzkpbfsyst $}}

\newcommand{\fzkpeffavg}   {\mbox{$11.3$}}
\newcommand{\kpeffavg}   {\mbox{$9.9$}}

\newcommand{\kcbf}       {\mbox{$4.6\pm1.0\pm0.4$}}
\newcommand{\kcfl}       {\mbox{$0.78\pm0.12\pm0.03$}}
\newcommand{\kcup}       {\mbox{$6.0$}}
\newcommand{\kcsig}      {\mbox{$5.3$}}

\newcommand{\fzkcbf}       {\mbox{$4.2\pm0.6\pm0.3$}}

\newcommand{\fzkcsig}      {\mbox{$9.0$}}

\newcommand{\kcAcp}       {\mbox{$0.31\pm0.13\pm0.03$}}
\newcommand{\kzAcp}       {\mbox{$0.25\pm0.14\pm0.03$}}
\newcommand{\kpAcp}       {\mbox{$0.59\pm0.31\pm0.03$}}

\newcommand{\fzkcAcp}       {\mbox{$-0.15\pm0.12\pm0.03$}}
\newcommand{\fzkzAcp}       {\mbox{$-0.34\pm0.16\pm0.03$}}
\newcommand{\fzkpAcp}       {\mbox{$0.14\pm0.12\pm0.03$}}

\begin{document}

\preprint{\babar-PUB-10/026}
\preprint{SLAC-PUB-14334}


\title{
\large \bfseries \boldmath 
Measurements of branching fractions, polarizations, and direct
\CP-violation asymmetries in 
\BtoRhozKstarp\ and \BtofzKstarp\ decays
}

\date{\today}
%
\author{P.~del~Amo~Sanchez}
\author{J.~P.~Lees}
\author{V.~Poireau}
\author{E.~Prencipe}
\author{V.~Tisserand}
\affiliation{Laboratoire d'Annecy-le-Vieux de Physique des Particules (LAPP), Universit\'e de Savoie, CNRS/IN2P3,  F-74941 Annecy-Le-Vieux, France}
\author{J.~Garra~Tico}
\author{E.~Grauges}
\affiliation{Universitat de Barcelona, Facultat de Fisica, Departament ECM, E-08028 Barcelona, Spain }
\author{M.~Martinelli$^{ab}$}
\author{D.~A.~Milanes}
\author{A.~Palano$^{ab}$ }
\author{M.~Pappagallo$^{ab}$ }
\affiliation{INFN Sezione di Bari$^{a}$; Dipartimento di Fisica, Universit\`a di Bari$^{b}$, I-70126 Bari, Italy }
% ...
\collaboration{The \babar\ Collaboration}
\noaffiliation

\begin{abstract}
We present measurements of the branching fractions, longitudinal
polarization, and direct \CP-violation asymmetries for the decays
\BtoRhozKstarp\ and \BtofzKstarp\ with a sample of \nbb\ \BB\ pairs
collected with the \babar\ detector at the \pep2\ asymmetric-energy
\epem\ collider at the SLAC National Accelerator Laboratory.  We
observe \BtoRhozKstarp\ with a significance of \kcsig$\sigma$ and
measure the branching fraction ${\calB}(\BtoRhozKstarp) =
(\kcbf)\times 10^{-6}$, the longitudinal polarization \fl\ = \kcfl,
and the \CP-violation asymmetry \Acp = \kcAcp. We observe
\BtofzKstarp\ and measure the branching fraction
${\calB}(\BtofzKstarp)\times{\calB}(\fz\to\pip\pim) = (\fzkcbf)\times
10^{-6}$ and the \CP-violation asymmetry \Acp = \fzkcAcp.  The first
uncertainty quoted is statistical and the second is systematic.
\end{abstract}


\maketitle


The study of the branching fractions and angular distributions of
\Bmeson\ decays to hadronic final states without a charm quark probes
the dynamics of both the weak and strong interactions. It also plays an
important role in understanding \CP\ violation in the quark sector,
constraining the Cabibbo-Kobayashi-Maskawa (CKM) matrix
parameters~\cite{bib:ckm} and searching for evidence for physics
beyond the standard model~\cite{bib:Beneke06,cheng08}.

The charmless decays \BtoRhoKstar\ proceed through penguin loops and
tree processes (\BtoRhopKstarz\ is a pure penguin process) to two vector
particles (VV). QCD factorization models predict a large longitudinal
polarization fraction \fL\ (of order $(1-4m_{\rm{V}}^2/m_B^2)\sim 0.9$) for
VV decays~\cite{bib:prediction}. However, measurements of 
penguin-dominated VV decays give \fL\ as low as
$\sim0.5$~\cite{bib:phiKst2}.  Several attempts to understand the
values of \fL\ within or beyond the standard model have been
made~\cite{bib:theory1}. 

For the \BtoRhozKstarp\ branching fraction, Beneke, Rohrer and
Yang~\cite{bib:Beneke06} predict the \CP-averaged branching fraction 
to be $(4.5^{+1.5+3.0}_{-1.3-1.4}) \times
10^{-6}$, while Cheng and Yang~\cite{cheng08} quote
$(5.5^{+0.6+1.3}_{-0.5-2.5}) \times 10^{-6}$, both based on QCD
factorization. The 90\% confidence level (C.L.) upper limit
\BtoRhozKstarp\ branching fraction has been measured to be $<6.1
\times 10^{-6}$~\cite{bib:RhoKst}.


We report measurements of branching fractions, longitudinal polarizations, 
and direct \CP-violating asymmetries for the decay modes
\BtoRhozKstarp\ and \BtofzKstarp\, where \rhoz\ and \Kstarp\ refer to 
the $\rho^0(770)$ and $K^{*+}(892)$ resonances, respectively.
The analysis is based on a data sample of \nbb\ \BB\ pairs,
equivalent to an integrated luminosity of \onreslumi, collected
with the \babar\ detector at the \pep2\ asymmetric-energy \epem\
collider operated at the SLAC National Accelerator Laboratory. The
\epem\ center-of-mass (c.m.) energy is $\sqrt{s} = 10.58$\gev,
corresponding to the \FourS\-resonance mass (on-resonance data). In
addition, \offreslumi\ of data collected 40~\mev\ below the
\FourS\-resonance (off-resonance data) are used for background
studies. We assume equal production rates of \BpBm\ and \BzBzb\ mesons
 and charge-conjugate modes are implied throughout~\cite{bib:PDG}.
The \babar\ detector is described in detail in 
Ref.~\cite{bib:babar}.


The \BtoRhozKstarp\ and \BtofzKstarp\ candidates are reconstructed
through the decays of \rhoz\ or \fz\ $\to\pip\pim$, $\Kstarp \to
\KS\pip$ or $\Kstarp \to \Kp\piz$, with $\KS\to\pip\pim$ and
$\piz\to\gamma\gamma$. The differential decay rate for \BtoRhozKstarp, after
integrating over the angle between the decay planes of the vector
mesons, for which the acceptance is nearly uniform, is proportional to
\begin{equation}
\frac{1-f_L}{4}\sin^2\theta_{\Kstarp}\sin^2\theta_{\rhoz} + 
   f_L \cos^2\theta_{\Kstarp}\cos^2\theta_{\rhoz},
\label{eq:helicity}
\end{equation}

\noindent where $\theta_{\Kstarp}$ ($\theta_{\rhoz}$) is the
helicity angle of the \Kstarp(\rhoz), defined as the angle
between the daughter $K$ (\pip) momentum and the direction opposite to
the \Bmeson\ momentum in the \Kstarp(\rhoz) rest
frame~\cite{bib:polarization}. The direct
\CP-violating asymmetry \Acp\ is defined as
$\Acp = (\Gamma^- - \Gamma^+)/(\Gamma^- + \Gamma^+)$, 
where $\Gamma^{\pm} = \Gamma(\Bpm\to f^{\pm})$ is the decay width for
a given charged final state $f^{\pm}$.


We apply the same selection criteria for \rhoz\ and \fz\ candidates.
The charged particles from the \Kstarp\ and \rhoz\ decays are required
to have a transverse momentum relative to the beam axis greater than
0.05\gevc. The particles are identified as either charged pions or
kaons by measurement of the energy loss in the tracking detectors, the
number of photons recorded by the ring-imaging Cherenkov detector and
the corresponding Cherenkov angle. These measurements are combined
with information from the electromagnetic calorimeter and
the instrumented magnetic-flux return detector, where appropriate, to
reject electrons, muons, and protons.

The \KS\ candidates are required to have a mass within 0.01\gevcc\ of
the nominal \KS\ mass~\cite{bib:PDG}, a decay vertex separated from
the \Bmeson\ decay vertex by at least 20 times the uncertainty in the
measurement of the separation of the vertex positions, a flight
distance in the direction transverse to the beam axis of at least
0.3\cm, and the cosine of the angle between the line joining the $B$
and \KS\ decay vertices and the \KS\ momentum greater than 0.999.

In the laboratory frame, the energy of each photon from the \piz\
candidate must be greater than 0.03\gev, the \piz\ energy must
exceed 0.25\gev, and the reconstructed \piz\ invariant mass is
required to be in the range $0.12\le m_{\gamma\gamma} \le 0.15$\gevcc.
After selection, the \piz\ candidate's mass is constrained to its
nominal value~\cite{bib:PDG}.

We require the invariant mass of the \Kstarp\ and \rhoz\ candidates to
satisfy $0.792 < \mkpi < 0.992$\gevcc and $0.52 < \mpipi <
1.05$\gevcc, respectively. A \Bmeson\ candidate is formed from the
\Kstarp\ and \rhoz\ candidates, with the condition that the \Kstarp\
and \rhoz\ candidates originate from the interaction region and the
$\chi^2$ of the \Bmeson\ vertex fit is less than 100. We require that
there is at least one additional charged track in the event and create
a vertex for a second \Bmeson\ from all remaining charged
tracks and neutral clusters that are consistent with originating from the 
interaction region.

The \Bmeson\ candidates are characterized kinematically by the energy
difference $\DeltaE = E^*_B - \sqrt{s}/2$ and the beam
energy-substituted mass $m_{\rm ES} = \left[ (s/2+{\bf p}\cdot{\bf
p}_B)^2/E^2-{\bf p}_B^2\right] ^{1/2}$, where $(E,{\bf p})$ and
$(E_B,{\bf p}_B)$ are the four-momenta of the \FourS and \Bmeson\
candidate in the laboratory frame, respectively, and the asterisk
denotes the c.m. frame.  The event sample is taken from the
region $|\DeltaE| <0.10$\gev and $5.255 \le \mes \le 5.289$\gevcc. The
extended \mes\ range ensures the shape of the background distribution
is properly modeled. Sideband events, outside the region $|\DeltaE|\le
0.07\gev$ and $5.270 \le \mes \le 5.289\gevcc$, are used to
characterize the background and cross-check the Monte Carlo (MC) 
background simulations~\cite{bib:geant}.

We suppress the background from \Bmesons\ decaying to charm by forming
the invariant mass $m_{D}$ from combinations of two or three out of the four
daughter particles' four-momenta. The event is rejected if $1.835 <
m_{D} < 1.895$\gevcc\ and the charge and particle type of the tracks
are consistent with a known decay from a
\Dmeson~\cite{bib:PDG}. Finally, to reduce the background and to avoid
the region where the reconstruction efficiency falls off rapidly for
low momentum tracks, we require the cosines of the helicity angles of
the \Kstarp\ and \rhoz\ candidates to satisfy $\cos\theta_{\Kstarp}
\le 0.92$ and $|\cos\theta_{\rhoz}| \le 0.95$, respectively.

To reject the background consisting of light-quark \qqbar $(q =
u,d,s,c)$ continuum events, we require $|\cos\theta_T|<0.85$, where
$\theta_T$ is the angle, in the c.m.\ frame, between the thrust axis
of the \Bmeson\ and that formed from the other tracks and neutral
clusters in the event. Signal events have an approximately uniform 
distribution in $|\cos\theta_T|$, while \qqbar\ continuum events peak
at $1$.

After the application of the selection criteria, the average number of
\RhozKstarp\ candidates per event with \KstarpKz\ in signal MC
simulations is 1.14 (1.03) for fully longitudinally (transversely)
polarized decays. The candidate with the smallest fitted decay vertex
$\chi^2$ is chosen. Up to 2.1\% (1.0\%) of longitudinally
(transversely) polarized MC signal events are misreconstructed, with
one or more tracks originating from the other \Bmeson\ in the event.
For \RhozKstarp\ with \KstarpKp, the average number of candidates per
event is 1.20 (1.08) and the fraction of misreconstructed candidates
is 5.9\% (2.7\%) for fully longitudinally (transversely) polarized
decays.  For \fzKstarp, the number of candidates per event and the
fraction of misreconstructed events are 1.02 (1.06) and 9.1\% (13.8\%)
for decays with \KstarpKz\ (\KstarpKp).  The \rhoz\ and \Kstarp\ masses and
widths in the MC simulation are taken from Ref.~\cite{bib:PDG} and we
use the measured \fz\ lineshape from Ref.~\cite{bib:e791}.

A neural net discriminant is used to provide additional separation
between signal and \qqbar\ continuum. It is constructed from six
variables calculated in the c.m. frame: the polar angles of the
\Bmeson\ momentum vector and the \Bmeson\ thrust axis with respect to
the beam axis, the angle between the \Bmeson\ thrust axis and the
thrust axis of the rest of the event, the ratio of the second- and
zeroth-order momentum-weighted polynomial moments of the energy flow
around the \Bmeson\ thrust axis~\cite{bib:Legendre}, the flavor of the
other \Bmeson\ as reported by a multivariate tagging
algorithm~\cite{bib:tagging}, and the boost-corrected proper-time
difference between the decays of the two \Bmesons\ divided by its
variance. The discriminant is trained using MC for the signal, and
\qqbar\ continuum MC and off-resonance data for the background.


We define an extended likelihood function to be used in an 
unbinned maximum likelihood (ML) fit as
\begin{equation}
{\mathcal L} = \frac{1}{N!}\exp{\left(-\sum_{j}n_{j}\right)}
\prod_{i=1}^N\left[\sum_{j}n_{j}{\mathcal
    P}_{j}(\vec{x}_i;\vec{\alpha}_j)\right]\!,
\end{equation}

\noindent where the likelihood ${\cal L}_i$ for each event
candidate $i$ is the sum of $n_j {\cal P}_j(\vec x_i; \vec \alpha_j)$
over hypotheses $j$: two signal modes \RhozKstarp and \fzKstarp\
(including misreconstructed signal candidates); \qqbar\ continuum
background; and nine \Bbacks\ as discussed below. ${\cal P}_j(\vec
x_i; \vec \alpha_j)$ is the product of the probability density
functions (PDFs) for hypothesis $j$ evaluated for the $i$-th event's
measured variables $\vec x_i$. The number of events for hypothesis $j$
is denoted by $n_j$ and $N$ is the total number of events in the
sample. The quantities $\vec \alpha_j$ represent parameters to
describe the expected distributions of the measured variables for each
hypothesis $j$.  Each discriminating variable $\vec x_i$ in the
likelihood function is modeled with a PDF, where the parameters $\vec
\alpha_j$ are extracted from MC simulation, off-resonance data, or
(\mes, \DeltaE) sideband data. The seven variables $\vec x_i$ used in
the fit are \mes, \DeltaE, the neural net output, \mpipi, \mkpi, 
the absolute cosine of the helicity angle of the \rhoz/\fz\
candidate and the cosine of the helicity of the \Kstarp\ 
candidate. Since most of the linear correlations among the fit
variable distributions are found to be about $1\%$, with a maximum of
11\%, we take each ${\cal P}_j$ to be the product of the PDFs for the
separate variables.

The decays \BtoDpiKz\ and \BtoDpiKp\ have large branching fractions
and a similar topology to the decays under consideration. They are
used as calibration channels. We apply the same selection criteria as
described above except that the neural net is trained on the MC
simulated data for the calibration channel under consideration; the
\DeltaE\ range is reduced to $|\DeltaE|<0.08$\gev; the \mkpi\ and
\mpipi\ mass criteria are replaced with a mass range $1.8445 <
m_{\Dzb} < 1.8845$\gevcc; and no \Dmeson\ veto is applied. We use the
selected data to verify that the ML fit is performing correctly and
that the MC is simulating the neural net, $\Delta E$ and $m_{\rm
ES}$ distributions.

Backgrounds from \BB\ decays involving charmed mesons are effectively
suppressed by applying the veto on the \Dmeson\ mass described above.
The \Bbacks\ that remain after the event selection criteria have been
applied are identified and modeled using MC simulation. We categorize
the \Bbacks\ in the ML fit into nine main groups.  Two groups
represent decays where either a \Kstarp\ or a \rhoz/\fz\ is
falsely reconstructed. 
Four groups represent nonresonant final states
\pippim\Kstarp, \rhoz\Kpip, \pippim\Kpip, and \fz\Kpip, where \Kpip\
stands for \KS\pip\ or \Kp\piz. The decays $\Bz\to\etapr\KS$ and
$\Bp\to\etapr\Kp$ peak at high $\cos\theta_{\Kstarp}$ and are assigned
their own category. We allow for decays from higher mass \KstarII\
states. All remaining \Bback\ decays that are not accounted for by the
above groups are assigned to a dominant remainder group.

The invariant mass distributions in the ML fit are modeled with
relativistic Breit--Wigner functions for the \Kstarp\ and \fz,
together with a polynomial of order up to four for the smoothly
varying distribution of misreconstructed candidates. Following
Ref.~\cite{bib:gs}, a modified relativistic Breit--Wigner function
is used for the \rhoz\ meson.
The \KstarII\ is modeled with the LASS parametrization,
which consists of the \KstarII\ resonance together with an
effective-range nonresonant component~\cite{bib:lass}.  For the
signal, the distributions of the cosine of the helicity angles are
described by Eq.~\ref{eq:helicity} multiplied by a polynomial
acceptance function that corrects for changes in efficiency as a
function of helicity angle. The correction also accounts for the
reduction in efficiency at helicity near $0.78$ introduced indirectly
by the criteria used to veto \Dmesons. For backgrounds, the cosine of
the helicity angle distribution is modeled with a polynomial. The
neural net distributions are modeled using either an empirical
nonparametric function~\cite{bib:nonparam} or a histogram.  For \mes,
an asymmetric Gaussian is used for the signal; the function
$x\sqrt{1-x^2}\exp[-\xi (1-x^2)]$ with $x=\mes/E^*_B$ and $\xi$ a free
parameter~\cite{bib:argus} is used for \qqbar\ continuum and \BB\
backgrounds; and a combination of an asymmetric Gaussian with a
polynomial is used for all other hypotheses.  For \DeltaE, two
Gaussians are used for signal and polynomials for all other
hypotheses.

We simultaneously fit for the branching fractions \calB, \Acp, and \fL\ (for
\BtoRhozKstarp\ only). 
We allow the yields for all hypotheses to float except
for \rhoz\Kpip\ and \fz\Kpip\ which are fixed to their predicted MC
yields, assuming a branching fraction of $1\times 10^{-6}$. The
predicted yields for the fixed modes are less than one event.
The PDF
parameters $\xi$ for \mes, the slope of the \DeltaE distribution, and
the polynomial coefficients and normalizations describing the mass and
helicity angle distributions are allowed to vary for the \qqbar\
continuum and \BB\ remainder groups. We validate the fitting procedure
and obtain the sizes of potential biases on the fit results by
applying the fit to ensembles of simulated experiments using the
extracted fitted yields from data.  The observed fit biases in the MC
samples are subtracted from the fitted yields measured in the data.


The results of the ML fits are summarized in Table~\ref{tab:summary},
where we assume a branching fraction of 100\% for \fz\to\pip\pim. 
For decays with \KstarpKz\ (\KstarpKp), the event sample is \kzntot\
(\kpntot), with \kznuds\ (\kpnuds) fitted \qqbar\ continuum events and
\kznbb\ (\kpnbb) events in the \Bback\ remainder group. The signal
significance $S$ is defined as $S=\sqrt{2\Delta\ln {\cal L}}$, where
$\Delta\ln {\cal L}$ is the change in log-likelihood from the maximum
value to the value when the number of signal events is set to zero, corrected for
systematic errors by convolving the likelihood function with a
Gaussian distribution with a variance equal to the total systematic
error defined below. The linear correlation coefficient between the
\RhozKstarp\ and \fzKstarp\ branching fractions is 0.25.

\begin{figure}[!ht]
\centerline{
\setlength{\epsfxsize}{0.5\linewidth}\leavevmode\epsfbox{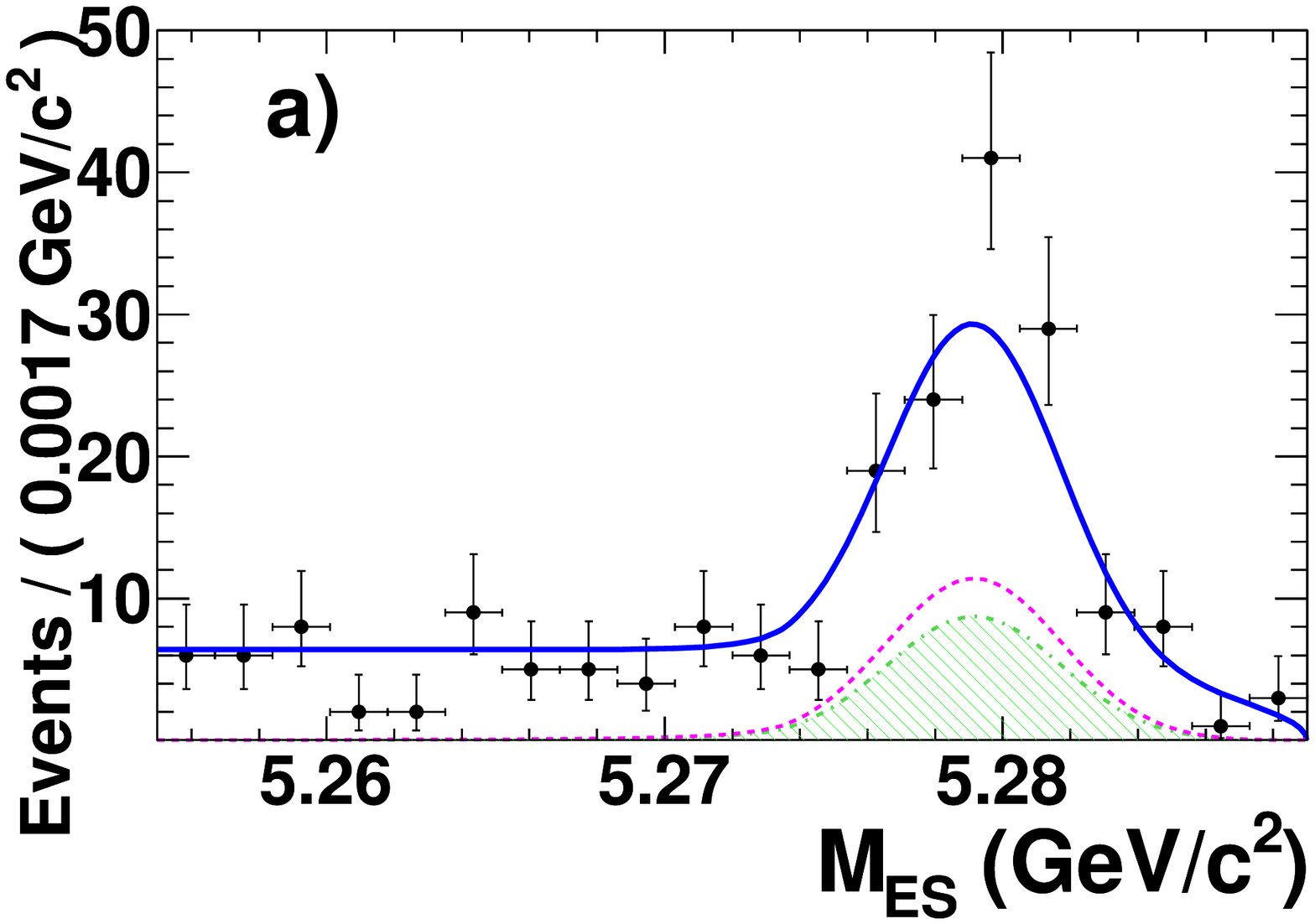}
\setlength{\epsfxsize}{0.5\linewidth}\leavevmode\epsfbox{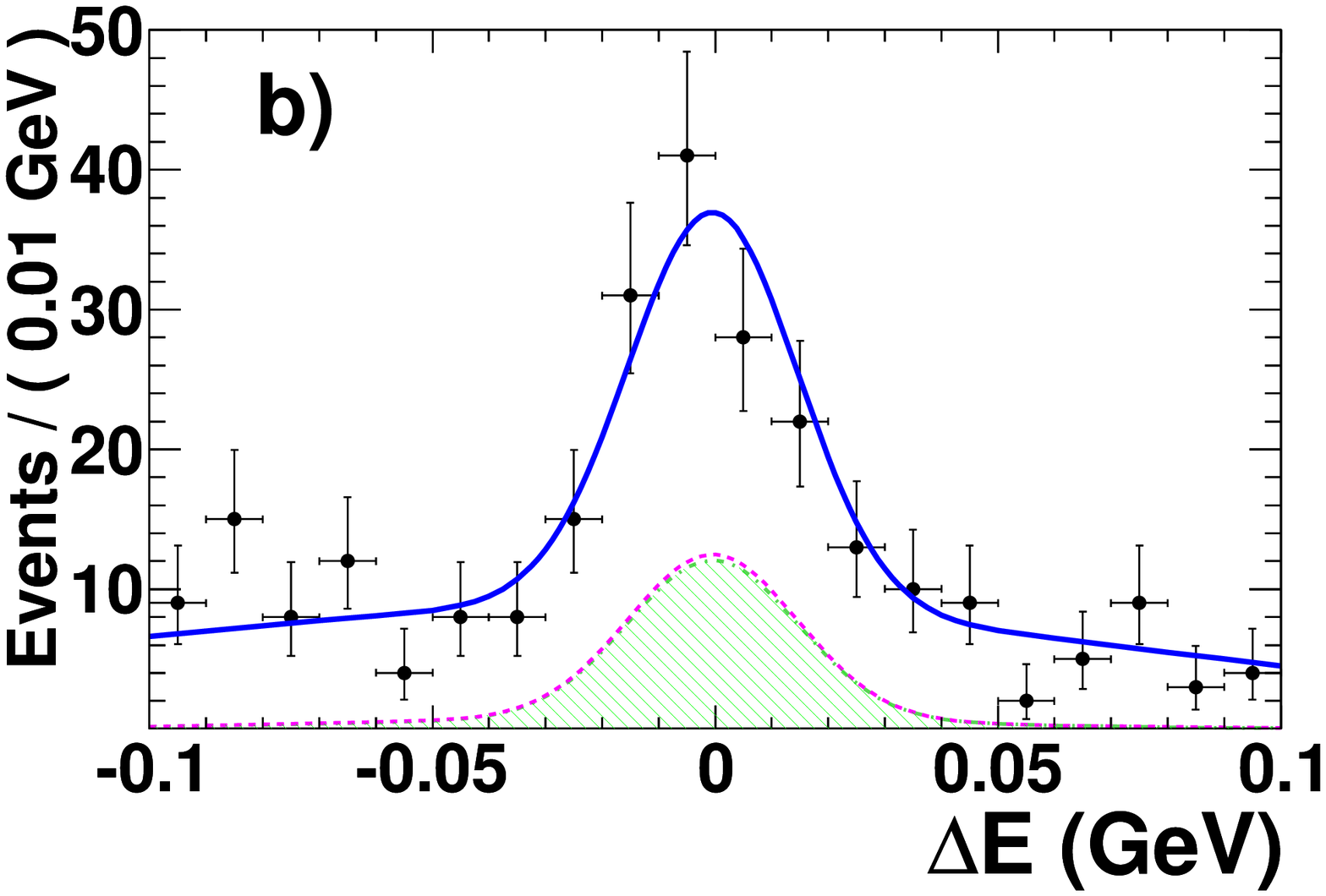}
}
\centerline{
\setlength{\epsfxsize}{0.5\linewidth}\leavevmode\epsfbox{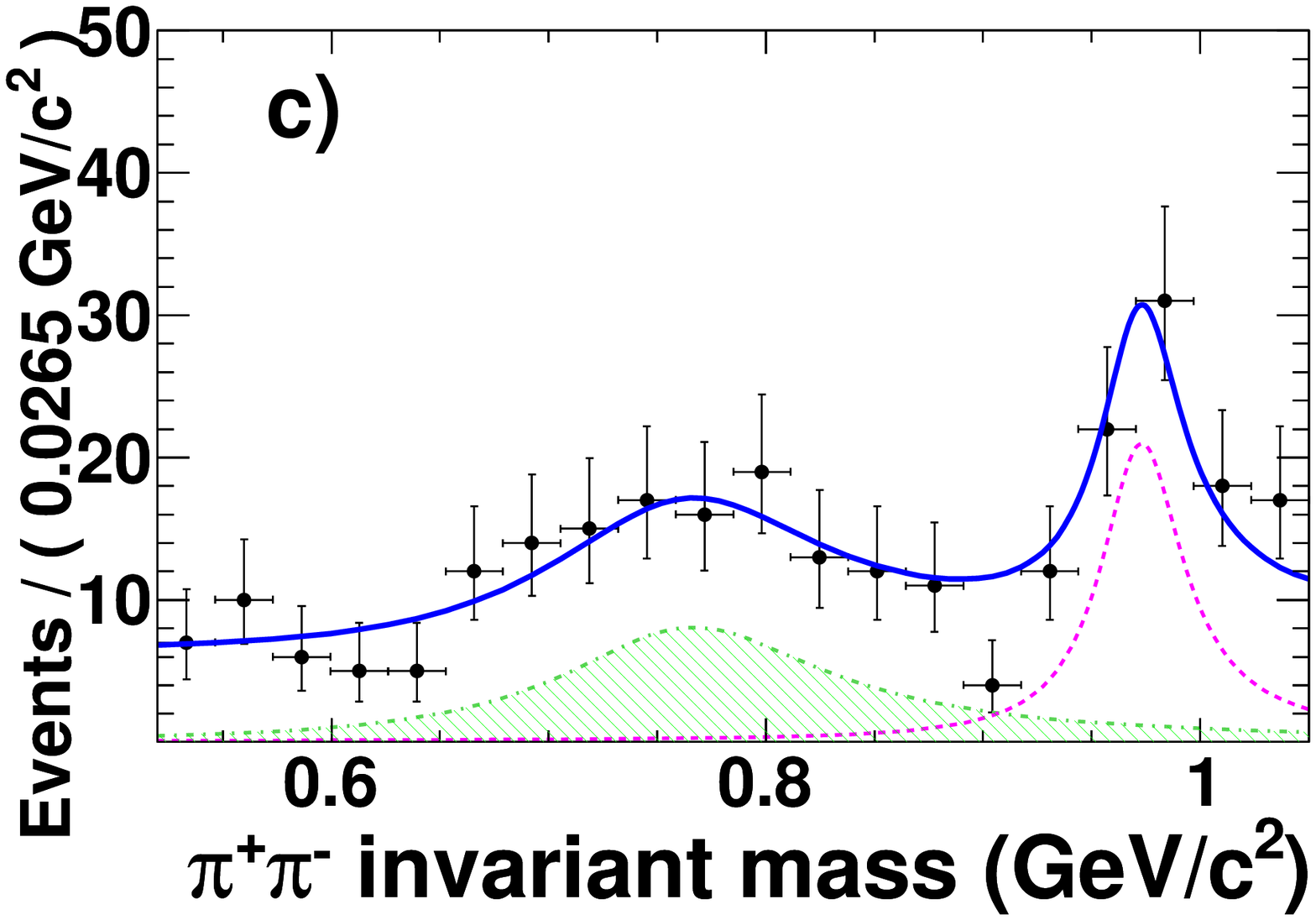}
\setlength{\epsfxsize}{0.5\linewidth}\leavevmode\epsfbox{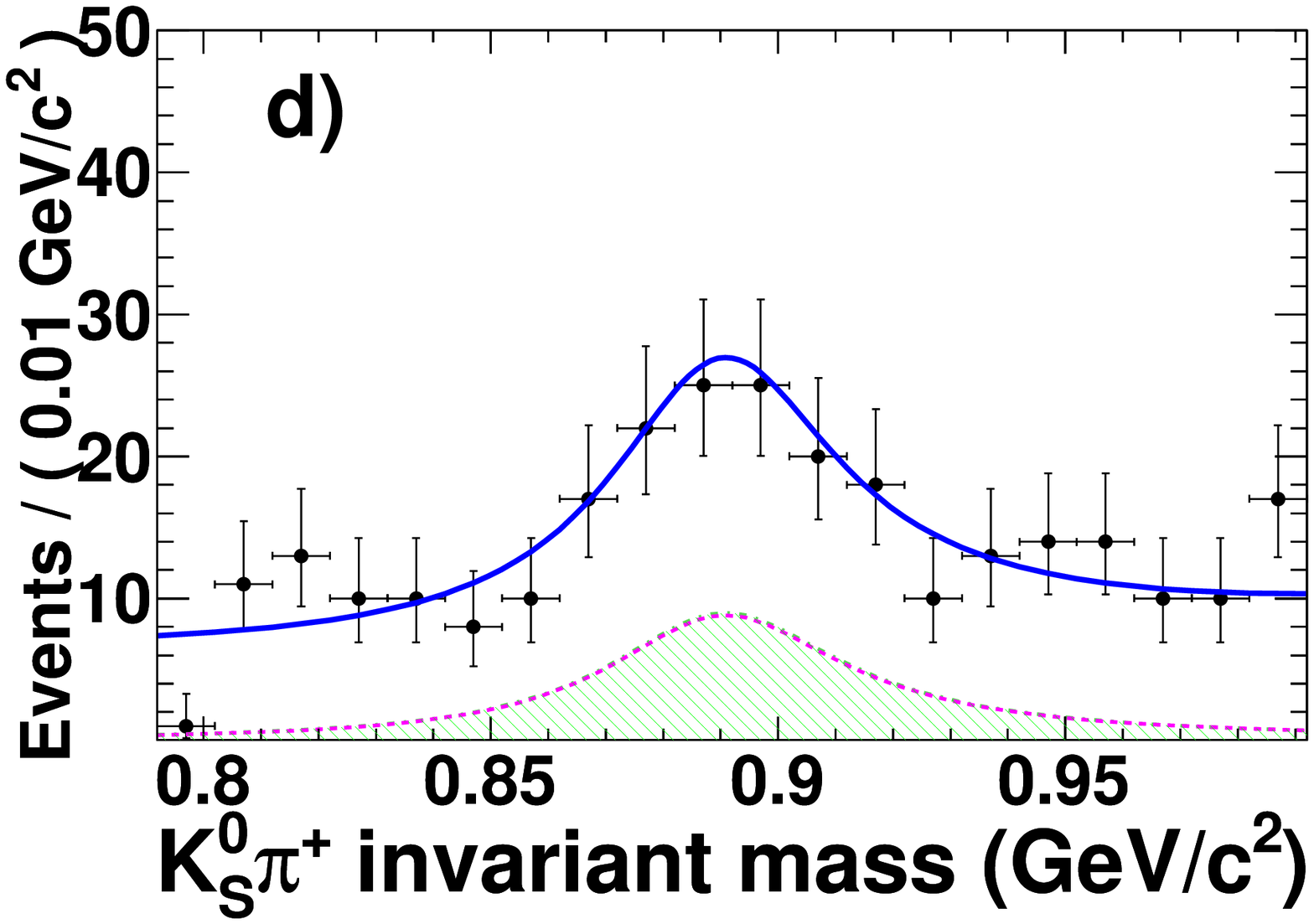}
}
\centerline{
\setlength{\epsfxsize}{0.5\linewidth}\leavevmode\epsfbox{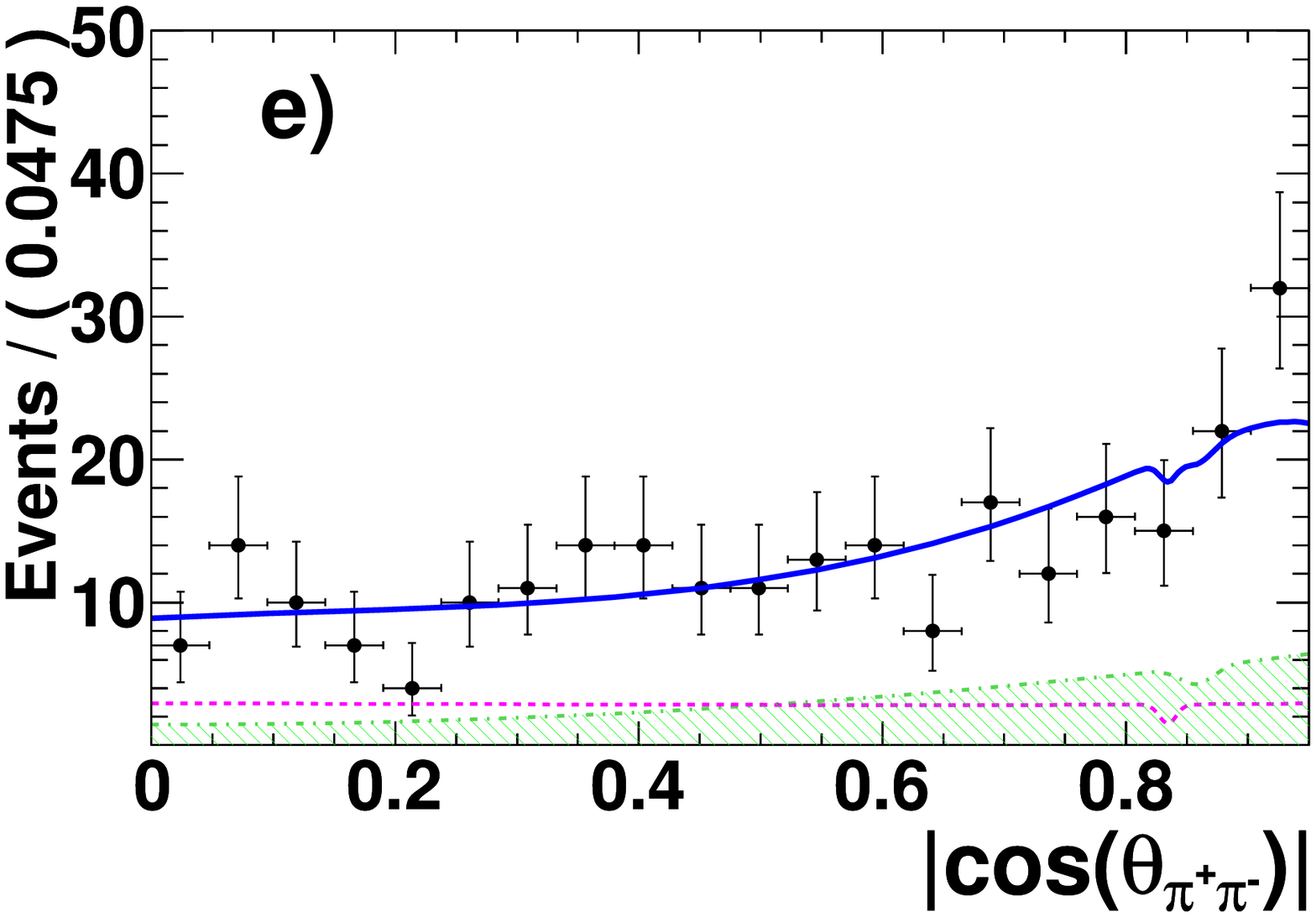}
\setlength{\epsfxsize}{0.5\linewidth}\leavevmode\epsfbox{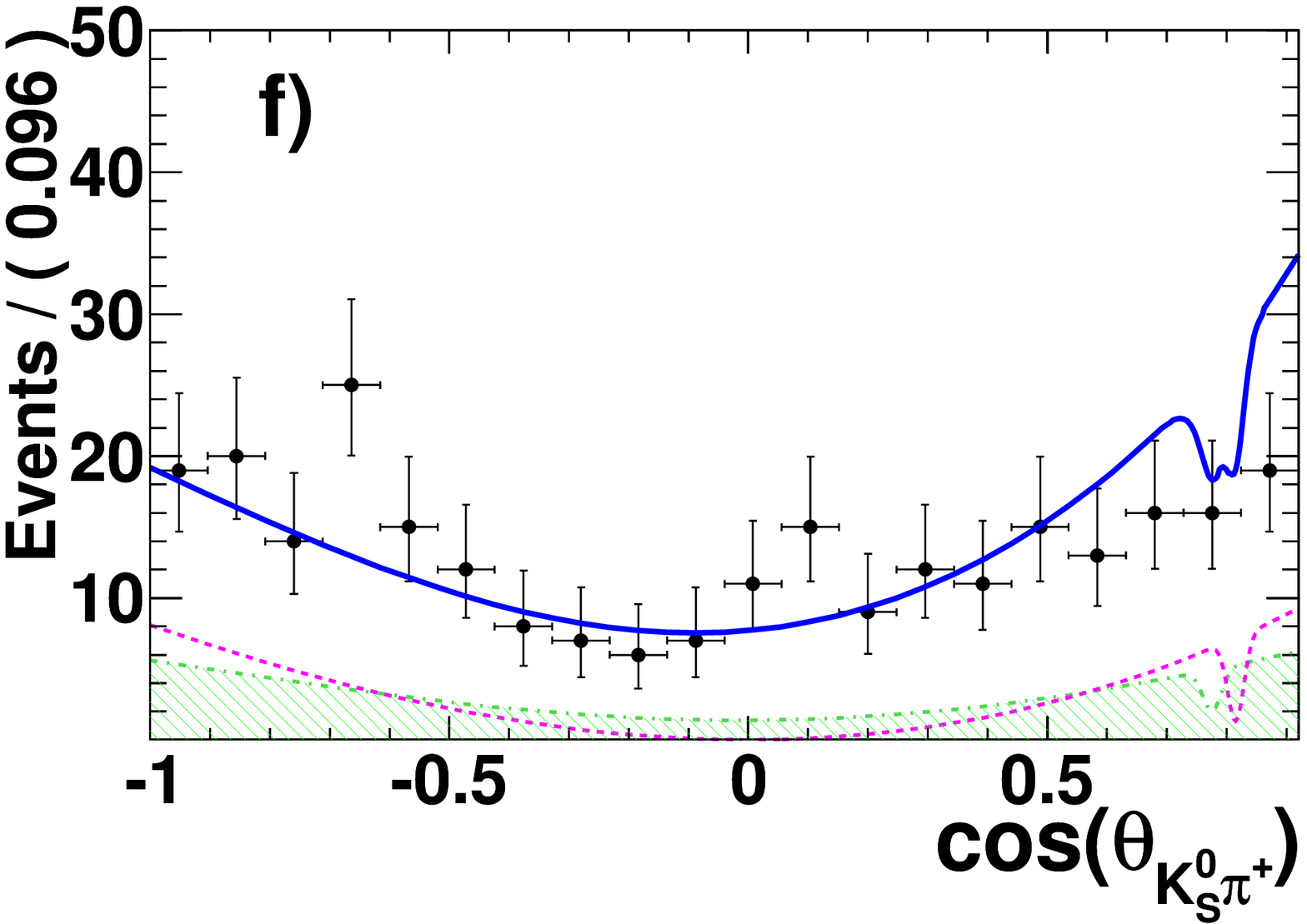}
}
\vspace{-0.3cm}
\caption{
\label{fig:proj1} Projections of the
multidimensional fit onto (a) \mes, (b) \DeltaE, (c) \pip\pim\ mass, (d) \KS\pip\ mass, (e)
 $|\cos\theta_{\pippim}|$, and (f) $\cos\theta_{\KS\pip}$ 
for modes with \Kstarp\to\KS\pip. The points with
 error bars show the data; the solid line shows
 signal-plus-background; the green hatched area is the \RhozKstarp\
 signal; and the red dashed line is the \fzKstarp\ signal.}
\label{fig:fig01}
\end{figure}

\begin{figure}[!ht]
\centerline{
\setlength{\epsfxsize}{0.5\linewidth}\leavevmode\epsfbox{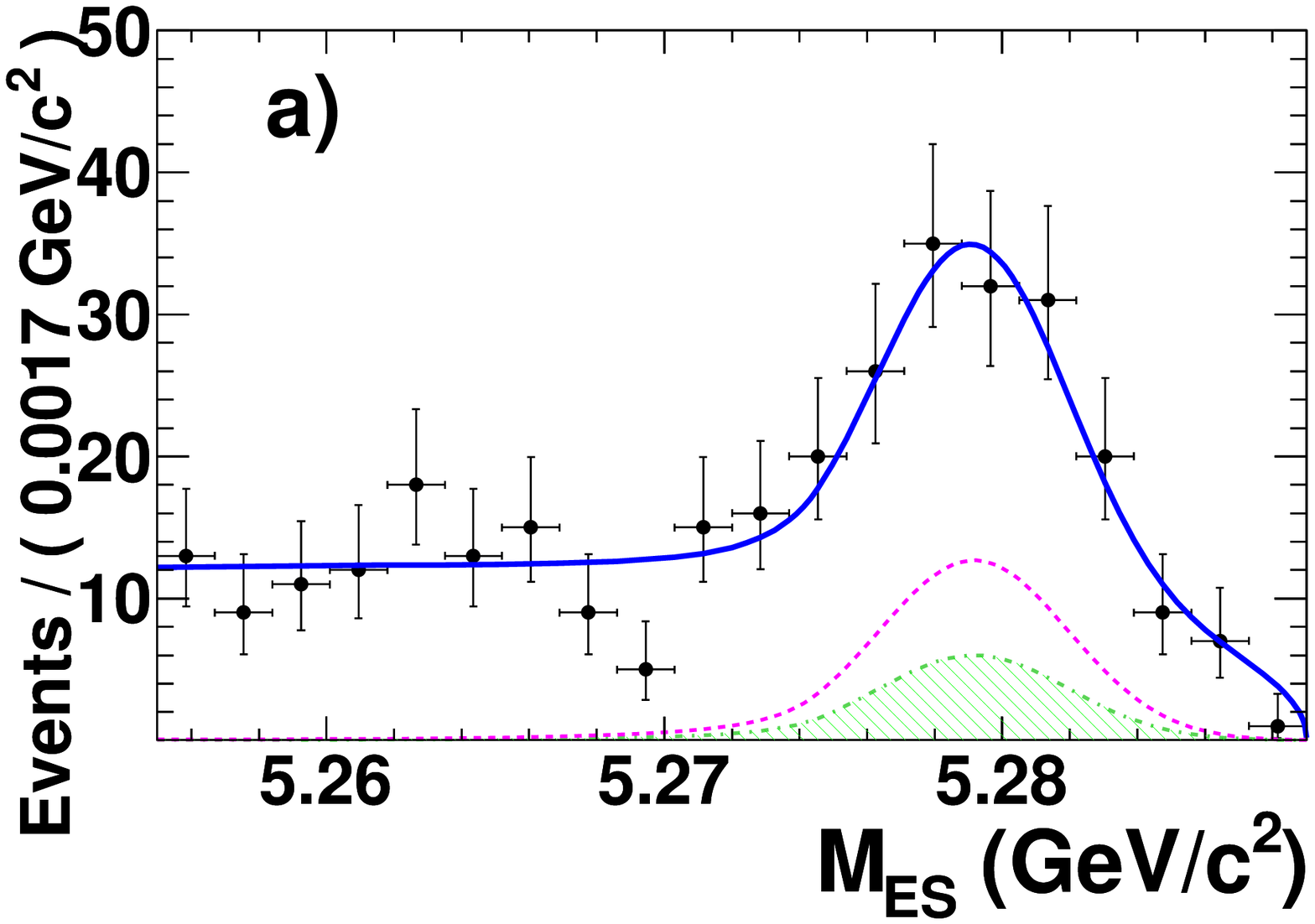}
\setlength{\epsfxsize}{0.5\linewidth}\leavevmode\epsfbox{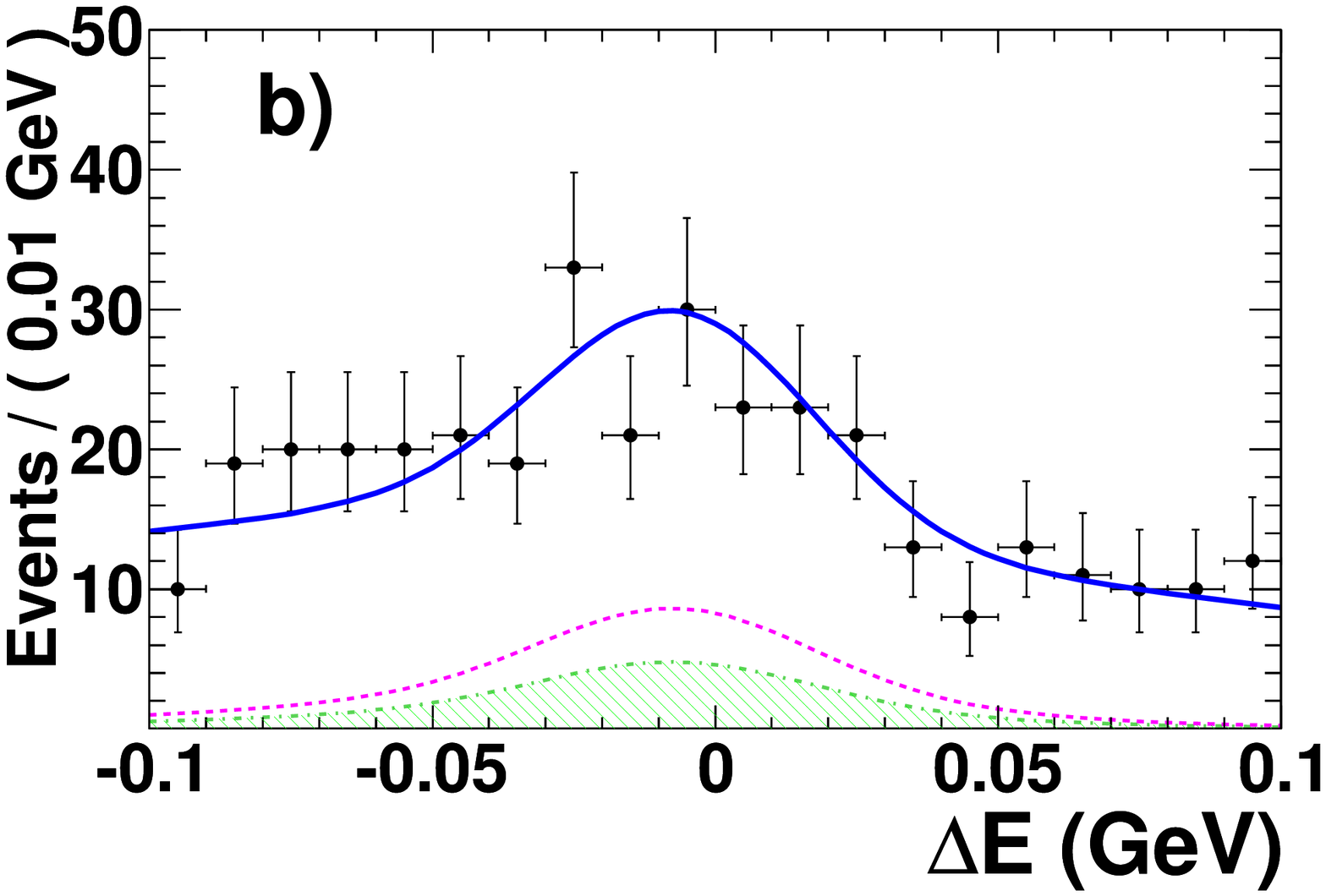}
}
\centerline{
\setlength{\epsfxsize}{0.5\linewidth}\leavevmode\epsfbox{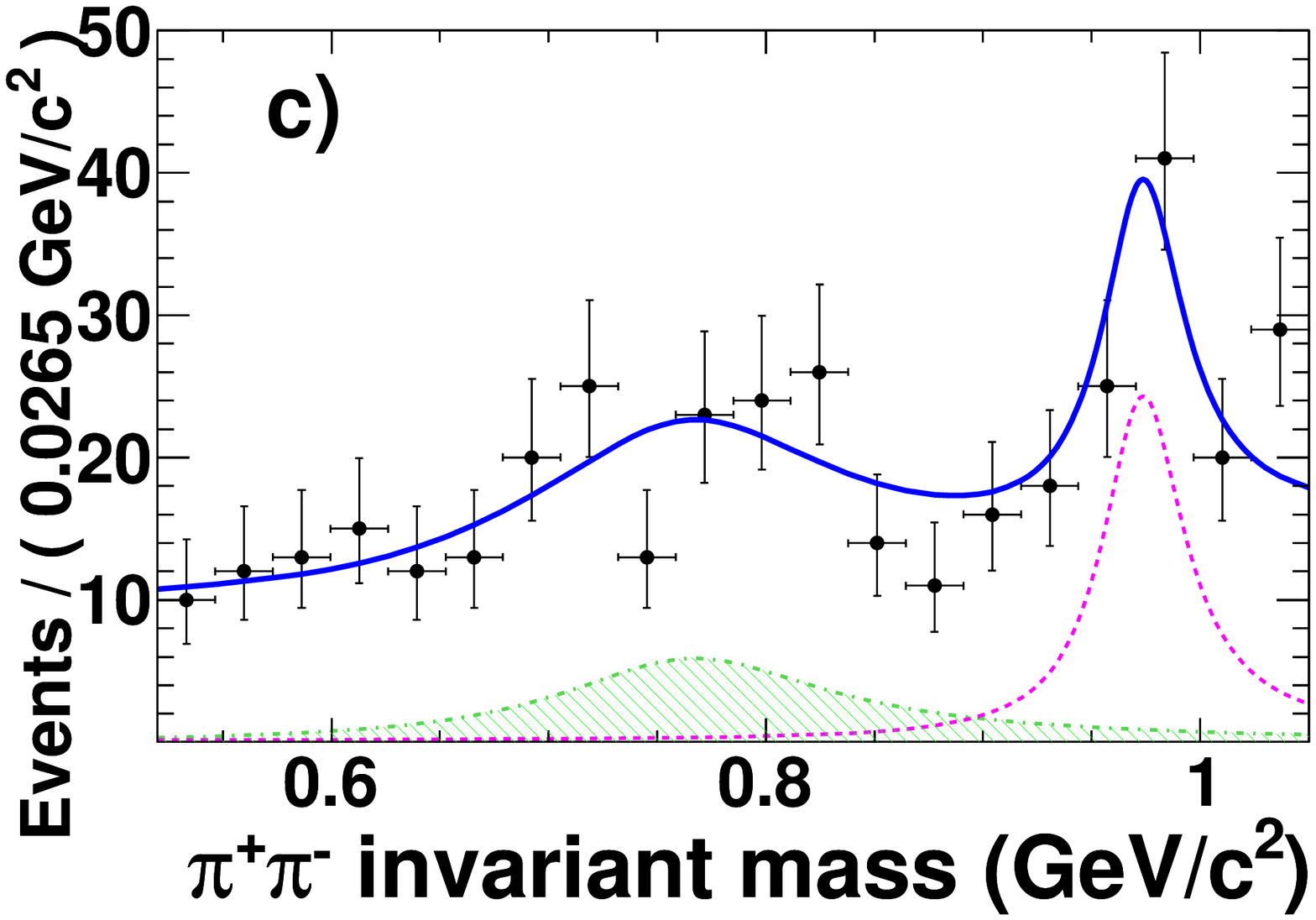}
\setlength{\epsfxsize}{0.5\linewidth}\leavevmode\epsfbox{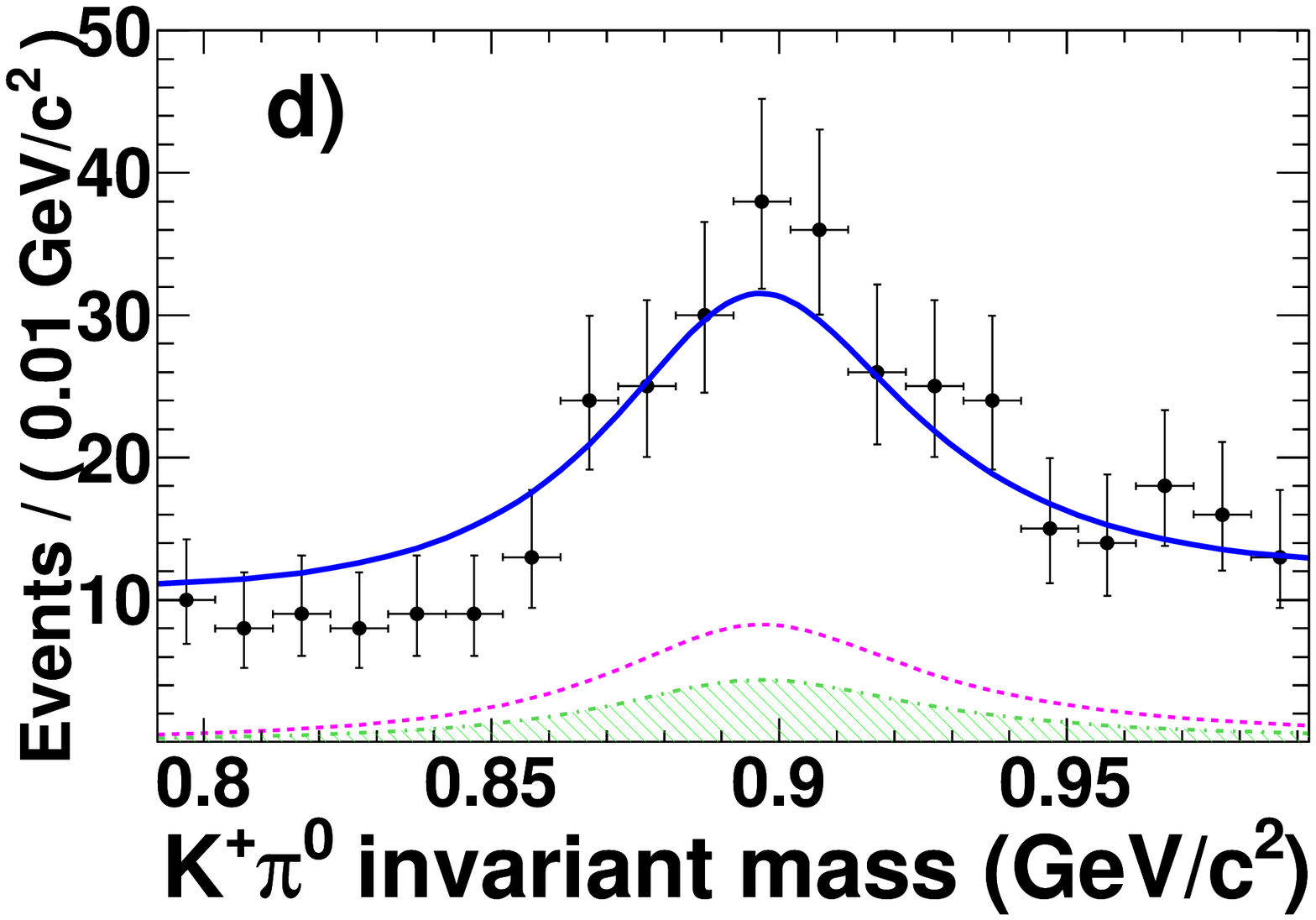}
}
\centerline{
\setlength{\epsfxsize}{0.5\linewidth}\leavevmode\epsfbox{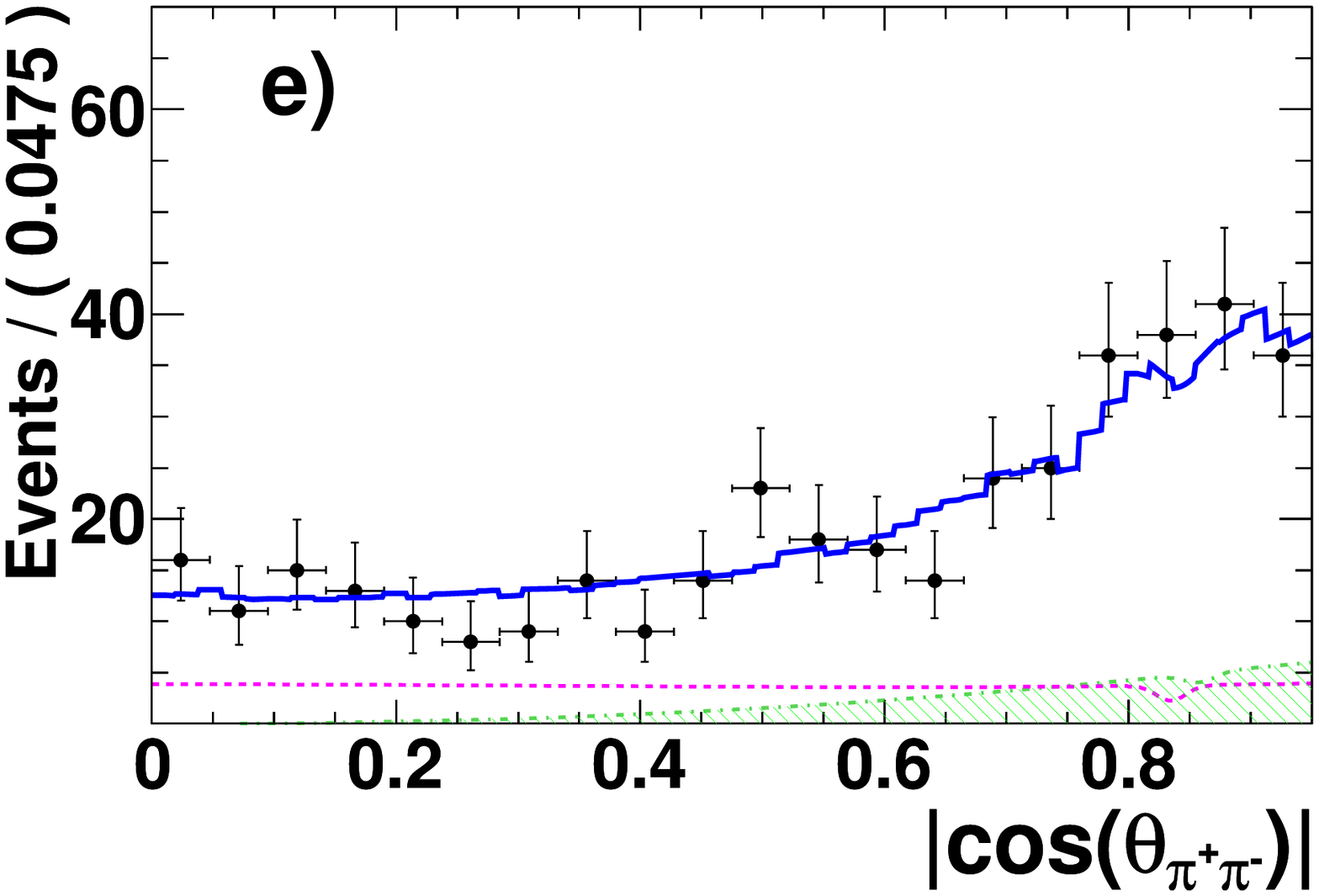}
\setlength{\epsfxsize}{0.5\linewidth}\leavevmode\epsfbox{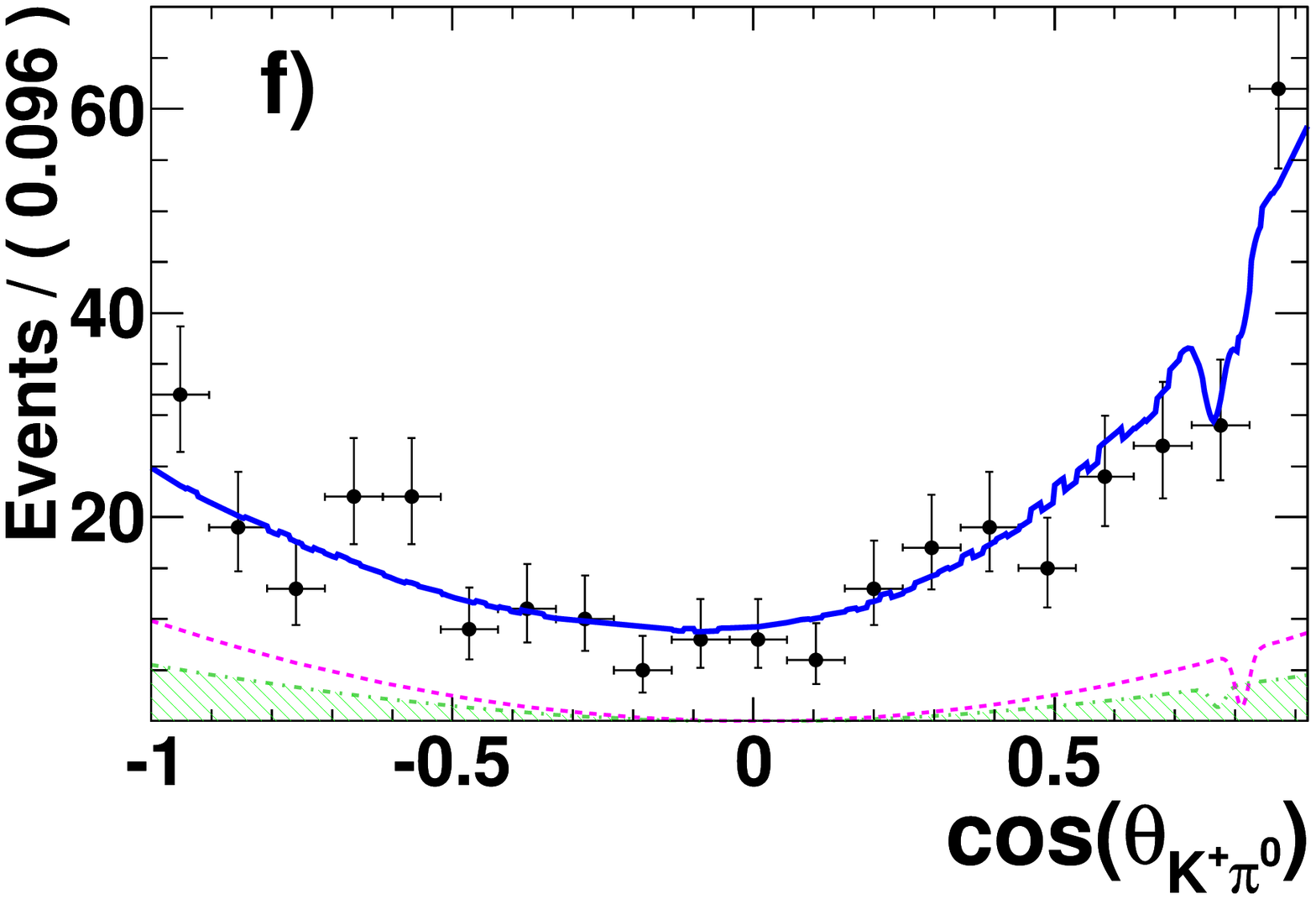}
}
\vspace{-0.3cm}
\caption{
\label{fig:proj2} Projections of the
multidimensional fit onto (a) \mes, (b) \DeltaE, (c) \pip\pim\ mass, (d) \Kp\piz\ mass, (e)
 $|\cos\theta_{\pippim}|$, and (c) $\cos\theta_{\Kp\piz}$ for modes
 with \Kstarp\to\Kp\piz. The same projection criteria and legend are
 used as in Fig.~\ref{fig:fig01}.}
\label{fig:fig02}
\end{figure}

Figs.~\ref{fig:proj1} and~\ref{fig:proj2} show the projections of the
fits onto \mes, \DeltaE, the masses, and the cosines of the helicity
angles for decays with \KstarpKz\ and \KstarpKp, respectively. The
candidates in the figures are subject to a requirement on the
probability ratio ${\cal P}_{\rm sig}/({\cal P}_{\rm sig} +{\cal
P}_{\rm bkg})>0.9$, where ${\cal P}_{\rm sig}$ and ${\cal P}_{\rm
bkg}$ are the signal and total background probabilities,
respectively, computed without the use of the variable plotted.

\begin{table*}[htbp!]
\caption[Summary of results] {Results for the measured \B decays:
  signal yield $Y$ (corrected for fit bias) and its statistical uncertainty, reconstruction
  efficiency (\%), daughter branching fraction product $\Pi \calB_i
  (\%)$~\cite{bib:PDG}, significance $S$ (with statistical and
  systematic uncertainties included), branching fraction \calB, 90\%
  C.L. upper limit (for modes with $S<6\sigma$), longitudinal
  polarization \fL\ and \CP-violating asymmetry \Acp.}
\begin{center}
\resizebox{1.0\textwidth}{!}{
\begin{tabular}{lcccccccc}
\hline\hline
\noalign{\vskip1pt}
Mode & $Y$ & $\epsilon$(\%) & $\Pi\calB_i(\%)$ & S($\sigma$) & \calB ($\times 10^{-6}$) & UL ($\times 10^{-6}$) & \fL & \Acp \\ \hline
\BtoRhozKstarp      & &           &      & \kcsig & \kcbf & \kcup & \kcfl & \kcAcp \\ 
\,\,\,\Kstarp\to\KS\pip & \kznsig & \kzeffavg & 23.1 & \kzsigfull & \kzbf & \kzup & \kzfl & \kzAcp \\ 
\,\,\,\Kstarp\to\Kp\piz & \kpnsig & \kpeffavg & 33.3 & \kpsigfull & \kpbf & \kpup & \kpfl & \kpAcp \\ 
\hline
\BtofzKstarp      & &             &      & \fzkcsig & \fzkcbf & - & - & \fzkcAcp \\ 
\,\,\,\Kstarp\to\KS\pip & \fzkznsig & \fzkzeffavg & 23.1 & \fzkzsigfull & \fzkzbf & - & - & \fzkzAcp \\ 
\,\,\,\Kstarp\to\Kp\piz & \fzkpnsig & \fzkpeffavg & 33.3 & \fzkpsigfull & \fzkpbf & - & - & \fzkpAcp \\ 
\hline\hline
\end{tabular}
}
\end{center}
\label{tab:summary}
\end{table*}

The systematic errors on the yields and branching fractions arise from
the PDFs, fit biases, \fz\ parameters, interference, \Bback\ yields,
and efficiencies. The PDF uncertainties are calculated by varying the
PDF parameters that are held fixed in the original fit by their
errors, taking into account correlations. The uncertainty from the fit
bias includes its statistical uncertainty from the simulated
experiments and half of the correction itself, added in quadrature.
We allow for uncertainties in the \fz\ lineshape by performing a
separate fit with the \fz\ mean and width as additional free
parameters. The effect of possible interference between the \fz\ and
\rhoz\ is estimated by adding the \fz\ and \rhoz\ amplitudes together
with a varying phase difference and using half the maximum change in
the yield as an uncertainty.  We test for the presence of a scalar
$f_0(600)$ (or $\sigma$) by adding it to our model, using the mass and
width reported in Ref.~\cite{bib:bugg}. The contribution of the
\Bbacks\ to the error is calculated by performing an ensemble of fits
to the data where backgrounds are either removed from the fit (for
those categories with a fitted number of events consistent with zero),
allowed to float (for the fixed backgrounds) or fixed to the expected
number of events calculated from MC. The error is calculated as half
the difference between the default fit and the maximum deviation seen
in the ensemble of fits.  Finally, the uncertainty on the longitudinal
polarization affects the calculated yield efficiency. All these errors
are additive in nature and affect the significance of the branching
fraction results. We assume the sources of the uncertainties that
contribute to the additive errors are uncorrelated when combined to
form the overall branching fractions. The PDF parameter uncertainty
contributes up to 0.4 signal events to the systematic error and the
fit bias between 2.4 and 0.8 events, depending on the signal mode. We
see no evidence for the $f_0(600)$ state.  The \fz\ lineshape and
interference account for up to 0.8 and 2.0 events, respectively. The
overall systematic error is dominated by the uncertainty in the
\Bbacks\ and, for \RhozKstarp, the systematic error on \fL. The total
additive systematic error on the \BtoRhozKstarp\ signal yield is 9.4
and 6.7 events for \Kstarp\to\KS\pip and \Kstarp\to\Kp\piz,
respectively, and for \BtofzKstarp\ it is 4.4 and 1.3 events,
respectively.

Multiplicative uncertainties 
include reconstruction efficiency uncertainties from tracking (0.8\%
per track added linearly), charged particle identification (1.1\% per
track added linearly), \piz\ identification (3.0\%), \KS\
identification (1.0\%), track multiplicity (1.0\%), the number of \BB\
pairs (1.1\%), and MC signal statistics (0.2\%).  The total
multiplicative branching fraction systematic error is 4.5\% and 5.3\%
for decays with \KstarpKz\ and \KstarpKp, respectively. The
multiplicative uncertainties for both sub-modes are correlated. The majority
of the systematic uncertainties on \fL\ and \Acp\ cancel and the error
is dominated by the uncertainty on the PDF parameters (0.02). The
uncertainty due to the dependence of the reconstruction efficiency on
the charge of the kaon is estimated from MC to be 0.005. The total
systematic is calculated to be $\pm0.03$ for all modes.


In summary, we observe \BtoRhozKstarp\ with a significance of
\kcsig$\sigma$.  We measure the branching fraction
${\calB}(\BtoRhozKstarp) = (\kcbf)\times 10^{-6}$, the longitudinal
polarization \fl\ = \kcfl, and the \CP-violating asymmetry \Acp = \kcAcp.  We
observe \BtofzKstarp\ and measure the branching fraction
${\calB}(\BtofzKstarp)\times{\calB}(\fz\to\pip\pim) = (\fzkcbf)\times
10^{-6}$ and the \CP-violating asymmetry \Acp = \fzkcAcp.  The
\BtoRhozKstarp\ branching fraction is compatible with theoretical
predictions~\cite{bib:Beneke06,cheng08}.

We are grateful for the excellent luminosity and machine conditions
provided by our \pep2\ colleagues, 
and for the substantial dedicated effort from
the computing organizations that support \babar.
The collaborating institutions wish to thank 
SLAC for its support and kind hospitality. 
This work is supported by
DOE
and NSF (USA),
NSERC (Canada),
CEA and
CNRS-IN2P3
(France),
BMBF and DFG
(Germany),
INFN (Italy),
FOM (The Netherlands),
NFR (Norway),
MES (Russia),
MICIIN (Spain),
STFC (United Kingdom). 
Individuals have received support from the
Marie Curie EIF (European Union),
the A.~P.~Sloan Foundation (USA)
and the Binational Science Foundation (USA-Israel).


\end{document}